\documentclass[a4paper,fleqn]{article}
\usepackage{amsmath}
\usepackage[dvips]{graphicx}
\begin{document}

\title{\textbf{Transformation and integrability of a generalized short pulse equation}}

\author{\textsc{Sergei Sakovich}\bigskip \\
\small Institute of Physics, National Academy of Sciences of Belarus \\
\small sergsako@gmail.com}

\date{}

\maketitle

\begin{abstract}
By means of transformations to nonlinear Klein--Gordon equations, we show that a generalized short pulse equation is integrable in two (and, most probably, only two) distinct cases of its coefficients. The first case is the original short pulse equation (SPE). The second case, which we call the single-cycle pulse equation (SCPE), is a previously overlooked scalar reduction of a known integrable system of coupled SPEs. We get the Lax pair and bi-Hamiltonian structure for the SCPE and show that the smooth envelope soliton of the SCPE can be as short as only one cycle of its carrier frequency.
\end{abstract}

\section{Introduction}

In this paper, we study the integrability of the nonlinear wave equation
\begin{equation}
u_{xt} = u + a u^2 u_{xx} + b u u_x^2 , \label{e1}
\end{equation}
where $a$ and $b$ are arbitrary constants, not equal zero simultaneously. The values of $a$ and $b$ change under the scale transformations of $u$, $x$ and $t$, but the ratio $a/b$ does not change in this way and serves as an essential parameter of \eqref{e1} therefore. This nonlinear equation \eqref{e1} is a slight generalization of the well-known integrable short pulse equation (SPE)
\begin{equation}
u_{xt} = u + \frac{1}{6} \left( u^3 \right)_{xx} \label{e2}
\end{equation}
which, in its turn, corresponds to the case of $a/b = 1/2$ in \eqref{e1}. The nonlinear equation \eqref{e2} appeared first in the context of differential geometry \cite{BRT,R}. Later the SPE \eqref{e2} was rediscovered in the context of nonlinear optics \cite{SW,CJSW}, in the problem of propagation of ultra-short infrared light pulses in silica optical fibers, and in this way it acquired its current name and significance. The SPE has been studied in many aspects, including its Lax pair \cite{BRT,R,SS1}, transformation to the sine-Gordon equation \cite{SS1,SS2,SS3}, recursion operator and hierarchy \cite{SS1,B1,B2}, bi-Hamiltonian structure and conserved quantities \cite{B1,B2}, soliton solutions and periodic solutions \cite{SS2,M1,M2,P1,P2}, wave breaking and well-posedness \cite{LPS,PS}, and integrable discretizations \cite{FMO}.

Our aim is to show that the generalized SPE \eqref{e1} is integrable in two (and, most probably, only two) distinct cases of its coefficients. The first case, with $a/b = 1/2$, is the original SPE \eqref{e2} up to a scale transformation of variables. The second case, with  $a/b = 1$, corresponds via a scale transformation of variables to the new nonlinear wave equation
\begin{equation}
u_{xt} = u + \frac{1}{2} u \left( u^2 \right)_{xx} \label{e3}
\end{equation}
which we call the single-cycle pulse equation (SCPE). We use this name because we show that the smooth envelope soliton of the SCPE \eqref{e3} can be as short as only one cycle of its carrier frequency. In Section~\ref{s2} of this paper, we transform the generalized SPE \eqref{e1} with any value of $a/b$ to a corresponding nonlinear Klein--Gordon equation whose nonlinearity depends on $a/b$. In Section~\ref{s3}, we use the previously known results on integrability of nonlinear Klein--Gordon equations and show in this way that the generalized SPE \eqref{e1} corresponds to integrable nonlinear Klein--Gordon equations in the cases of $a/b = 1/2$ and $a/b = 1$ only. Next we concentrate on the SCPE \eqref{e3}, reveal its relation to a known integrable system of coupled SPEs, obtain its Lax pair and bi-Hamiltonian structure, and study its soliton solutions. Section~\ref{s4} contains concluding remarks.

\section{Transformation} \label{s2}

Let us show how to transform the generalized SPE \eqref{e1} with any value of $a/b$ to a corresponding nonlinear Klein--Gordon equation.

In the case of $a=0$, we have $b \ne 0$ and make $b=1$ in \eqref{e1} by a scale transformation of variables, without loss of generality. Then it is easy to see that the new dependent variable $w(x,t)$,
\begin{equation}
w = \arctan u_x , \label{e4}
\end{equation}
satisfies the nonlinear Klein--Gordon equation
\begin{equation}
w_{xt} = \tan w \label{e5}
\end{equation}
if $u$ satisfies the considered case of the generalized SPE \eqref{e1},
\begin{equation}
u_{xt} = u + u u_x^2 . \label{e6}
\end{equation}
Note that the inverse transformation from \eqref{e5} to \eqref{e6},
\begin{equation}
u = w_t , \label{e7}
\end{equation}
is also a local transformation, that is, like \eqref{e4}, it requires no integration.

From now on, we consider the case of $a \ne 0$ and follow the way of transformation used in \cite{SS3}. Introducing the new independent variable $y$,
\begin{equation}
x = x(y,t) , \qquad u(x,t) = p(y,t) , \label{e8}
\end{equation}
and imposing the condition
\begin{equation}
x_t = - a p^2 \label{e9}
\end{equation}
on the function $x(y,t)$ to considerably simplify the result, we cast the studied equation \eqref{e1} into the form
\begin{equation}
x_y p_{yt} + (2a-b) p p_y^2 - p x_y^2 = 0 . \label{e10}
\end{equation}
Note that this equation \eqref{e10} is invariant under the transformation $y \mapsto Y(y)$ with any function $Y$. This means that solutions of the system of equations \eqref{e9} and \eqref{e10} determine solutions of the studied equation \eqref{e1} parametrically, with $y$ being the parameter. Next we introduce the new dependent variable $q(y,t)$, such that
\begin{equation}
x_y = \frac{1}{q} p_y , \label{e11}
\end{equation}
which means that $q(y,t) = u_x (x,t)$. Compatibility condition $x_{ty} = x_{yt}$ for \eqref{e9} and \eqref{e11} reads
\begin{equation}
p_{yt} = \frac{1}{q} p_y q_t - 2a p q p_y . \label{e12}
\end{equation}
Eliminating $x_y$ from \eqref{e10} and \eqref{e11}, and using \eqref{e12}, we obtain the expression for $p$ in terms of $q$,
\begin{equation}
p = \frac{q_t}{1 + b q^2} , \label{e13}
\end{equation}
and the third-order equation for $q$,
\begin{equation}
\left( \log \left[ \left( \frac{q_t}{1 + b q^2} \right)_y \right] \right)_t - \frac{q_t}{q} + \frac{2a q q_t}{1 + b q^2} = 0 . \label{e14}
\end{equation}
Solutions $q(y,t)$ of this equation \eqref{e14} determine solutions of the second-order equation \eqref{e1} parametrically, via \eqref{e8}, \eqref{e9}, \eqref{e11} and \eqref{e13}. The fact that the order of \eqref{e14} exceeds the order of \eqref{e1} by one (hence, there is one extra arbitrary function in the general solution of \eqref{e14}) means that the arbitrariness $y \mapsto Y(y)$ of the parameter $y$ is still not fixed.

Let $b=0$. Since $a \ne 0$, we make $a=1$ in \eqref{e1} by a scale transformation of variables, without loss of generality. In this case, integrating \eqref{e14} over $t$, we get
\begin{equation}
\log q_{yt} - \log q + q^2 = c(y) , \label{e15}
\end{equation}
where the arbitrary function $c(y)$ is the ``constant'' of integration. We choose $c(y)=0$ without loss of generality, because it is always possible to make $c(y)=0$ in \eqref{e15} by the transformation $y \mapsto Y(y)$ with a properly chosen function $Y$. Note that, when the function $c(y)$ is fixed, the arbitrariness of the parameter $y$ is reduced only to the shifts $y \mapsto y + y_0$ with any constant $y_0$. As the result, we obtain that solutions of the considered case of the generalized SPE \eqref{e1},
\begin{equation}
u_{xt} = u + u^2 u_{xx} , \label{e16}
\end{equation}
are determined parametrically by solutions of the nonlinear Klein--Gordon equation
\begin{equation}
q_{yt} = q \exp \left( - q^2 \right) \label{e17}
\end{equation}
via the relations
\begin{gather}
u(x,t) =q_t (y,t) , \notag \\
x = x(y,t) : \qquad x_y = \exp \left( - q^2 \right) , \qquad x_t = - q_t^2 , \label{e18}
\end{gather}
where $y$ serves as the parameter.

Let $b \ne 0$. In this case, we make $b=1$ in \eqref{e1} by a scale transformation of variables, integrate \eqref{e14} over $t$, and get
\begin{equation}
\log \left[ \left( \arctan q \right)_{yt} \right] - \log q + a \log \left( 1 + q^2 \right) = c(y) , \label{e19}
\end{equation}
where the arbitrary function $c(y)$ is the ``constant'' of integration. Next we make $c(y)=0$ in \eqref{e19} by the transformation $y \mapsto Y(y)$ with a properly chosen $Y(y)$, and introduce the new dependent variable $r(y,t)$,
\begin{equation}
r = \arctan q . \label{e20}
\end{equation}
As the result, we obtain that solutions of the considered case of the generalized SPE \eqref{e1},
\begin{equation}
u_{xt} = u + a u^2 u_{xx} + u u_x^2 , \label{e21}
\end{equation}
are determined parametrically by solutions of the nonlinear Klein--Gordon equation
\begin{equation}
r_{yt} = \sin r ( \cos r )^{2a-1} \label{e22}
\end{equation}
via the relations
\begin{gather}
u(x,t) = r_t (y,t) , \notag \\
x = x(y,t) : \qquad x_y = ( \cos r )^{2a} , \qquad x_t = - a r_t^2 , \label{e23}
\end{gather}
where $y$ serves as the parameter. Note that $a$ is an arbitrary nonzero constant in this case. However, if we set $a=0$, the expressions \eqref{e22} and \eqref{e23} correctly reproduce the expressions \eqref{e5} and \eqref{e7}, respectively.

\section{Integrability} \label{s3}

We have transformed the generalized SPE \eqref{e1} with any value of $a/b$ to a corresponding nonlinear Klein--Gordon equation whose nonlinearity depends on the value of $a/b$. Now, using previously known results on integrability of nonlinear Klein--Gordon equations, we can draw a conclusion on integrability of the generalized SPE.

Integrability of nonlinear Klein--Gordon equations has been studied very well. According to the classification made in \cite{ZS}, the equation
\begin{equation}
z_{\xi \eta} = f(z) \label{e24}
\end{equation}
possesses a higher symmetry if and only if the function $f(z)$ satisfies one of the following two conditions:
\begin{equation}
f' = \alpha f \label{e25}
\end{equation}
or
\begin{equation}
f'' = \alpha f + \beta f' , \label{e26}
\end{equation}
where $z = z( \xi , \eta )$, the prime denotes the derivative with respect to $z$, the constant $\alpha$ in \eqref{e25} is arbitrary, while the constants $\alpha$ and $\beta$ in \eqref{e26} must satisfy the condition
\begin{equation}
\beta \left( \alpha - 2 \beta^2 \right) = 0 . \label{e27}
\end{equation}
Consequently, up to scalings and shifts of variables, only three distinct nonlinear equations of the form \eqref{e24} possess nontrivial groups of higher symmetries: the Liouville equation (Darboux integrable), the sine-Gordon equation (Lax integrable), and the Tzitzeica equation (Lax integrable). No more integrable nonlinear equations of the form \eqref{e24} have been discovered by various methods as yet.

The right-hand sides of the nonlinear Klein--Gordon equations \eqref{e5} and \eqref{e17} do not satisfy the conditions \eqref{e25} and \eqref{e26}. The right-hand side of the nonlinear Klein--Gordon equation \eqref{e22} fails the condition \eqref{e25} as well, but it satisfies the condition \eqref{e26} provided that $a=1/2$ or $a=1$. In the case of $a=1/2$, we obtain from \eqref{e21}--\eqref{e23} the well-known transformation \cite{SS2}
\begin{gather}
u(x,t) = r_t (y,t) , \notag \\
x = x(y,t) : \qquad x_y = \cos r , \qquad x_t = - \frac{1}{2} r_t^2 \label{e28}
\end{gather}
which relates the original SPE \eqref{e2} with the sine-Gordon equation
\begin{equation}
r_{yt} = \sin r . \label{e29}
\end{equation}
In the case of $a=1$, using the new dependent variable $s(y,t)$,
\begin{equation}
s = 2 r , \label{e30}
\end{equation}
we obtain from \eqref{e21}--\eqref{e23} the transformation
\begin{gather}
u(x,t) = \frac{1}{2} s_t (y,t) , \notag \\
x = x(y,t) : \qquad x_y = \frac{1}{2} + \frac{1}{2} \cos s , \qquad x_t = - \frac{1}{4} s_t^2 \label{e31}
\end{gather}
which relates the SCPE \eqref{e3} with the sine-Gordon equation, too,
\begin{equation}
s_{yt} = \sin s . \label{e32}
\end{equation}

Consequently, there are two (and, most probably, only two) distinct integrable cases of the generalized SPE \eqref{e1}, namely, the original SPE \eqref{e2} and the SCPE \eqref{e3}, and they are two different ``avatars'' of one and the same sine-Gordon equation. The words ``most probably'' mean, of course, that the validity of our conclusion relies on the completeness of the known list of integrable nonlinear Klein--Gordon equations.

From now on, we study the new integrable equation \eqref{e3}. Since we know the transformation \eqref{e31} relating the SCPE \eqref{e3} with the sine-Gordon equation \eqref{e32}, we can derive the Lax pair, bi-Hamiltonian structure and soliton solutions of the SCPE from the corresponding known objects of the sine-Gordon equation, in the way successfully used in \cite{SS2,S1,S2,BS1,BS2,BS3} for other equations. There is, however, the following easier way, at least for what concerns the Lax pair and bi-Hamiltonian structure.

Let us give one example of how our result on integrability of the generalized SPE \eqref{e1} can be used. Consider the system of two symmetrically coupled SPEs
\begin{equation}
u_{xt} = u + \frac{1}{6} \left( u^3 \right)_{xx} + g v^2 u_{xx} , \qquad v_{xt} = v + \frac{1}{6} \left( v^3 \right)_{xx} + g u^2 v_{xx} , \label{e33}
\end{equation}
where $g$ is an arbitrary constant. This system is a slight generalization of the integrable system of Feng \cite{F} which, in its turn, corresponds to the case of $g = 1/2$ in \eqref{e33}. Are there any other integrable cases of the system \eqref{e33} besides the known case with $g = 1/2$? If we set $v=0$ or $u=0$ in \eqref{e33}, this two-component system reduces to the integrable SPE \eqref{e2} for $u$ or $v$, respectively. However, if we set
\begin{equation}
v = \pm u , \label{e34}
\end{equation}
the system \eqref{e33} reduces to the generalized SPE
\begin{equation}
u_{xt} = u + \left( g + \frac{1}{2} \right) u^2 u_{xx} + u u_x^2 \label{e35}
\end{equation}
which, as we have already shown, is integrable in two (and, most probably, only two) cases. The case of $g=0$ in \eqref{e35}, when the equations in \eqref{e33} are decoupled, is the SPE \eqref{e2}. The case of $g = 1/2$ in \eqref{e35}, when \eqref{e33} is the system of Feng, is the SCPE \eqref{e3}. Taking into account that reductions of an integrable system must be integrable themselves, we conclude that the system of Feng is (most probably) the only integrable case of the coupled SPEs \eqref{e33}. As a by-product, we have established the fact which was surprisingly overlooked in the literature till now, namely, that the system of Feng \cite{F} possesses two different scalar reductions, the SPE \eqref{e2} and the SCPE \eqref{e3}.

Since the Lax pair and bi-Hamiltonian structure of the system of Feng have already been obtained in \cite{BS3}, we can use them to obtain the Lax pair and bi-Hamiltonian structure of the SCPE \eqref{e3} via the reduction \eqref{e34}. Taking from \cite{BS3}  the Lax pair of the system of Feng and setting $v = - u$ (note the choice of the sign), we get the following Lax pair of the SCPE \eqref{e3}:
\begin{equation}
\Psi_x = X \Psi , \qquad \Psi_t = T \Psi \label{e36}
\end{equation}
with
\begin{gather}
X =
\begin{pmatrix}
\lambda \left( 1 - u_x^2 \right) & 2 \lambda u_x \\
2 \lambda u_x & - \lambda \left( 1 - u_x^2 \right)
\end{pmatrix}
, \notag \\
T =
\begin{pmatrix}
\lambda u^2 \left( 1 - u_x^2 \right) + \frac{1}{4 \lambda} & 2 \lambda u^2 u_x - u \\
2 \lambda u^2 u_x + u & - \lambda u^2 \left( 1 - u_x^2 \right) - \frac{1}{4 \lambda}
\end{pmatrix}
, \label{e37}
\end{gather}
where $\Psi (x,t)$ is a two-component column, and $\lambda$ is the spectral parameter. The choice of $v=u$, however, would bring us to a ``fake'' Lax pair \eqref{e36} with some diagonal matrices $X$ and $T$, which is equivalent to the conservation law
\begin{equation}
\left( u_x^2 \right)_t + \left( - u^2 - u^2 u_x^2 \right)_x = 0 \label{e38}
\end{equation}
of the SCPE \eqref{e3}. Next, taking from \cite{BS3} the bi-Hamiltonian structure of the system of Feng and applying the reduction \eqref{e34} with any choice of the sign, we get the following bi-Hamiltonian structure of the SCPE \eqref{e3}:
\begin{equation}
D = \partial_x^{-1} , \qquad H = \int dx \left( \frac{1}{2} u^2 - \frac{1}{2} u^2 u_x^2 \right) , \label{e39}
\end{equation}
and
\begin{gather}
D = \left[ \partial_x^{-1} \left( 1 - u_x^2 \right) + 2 u_x \partial_x^{-1} u_x \right] \partial_x^{-1} \left[ \left( 1 - u_x^2 \right) \partial_x^{-1} + 2 u_x \partial_x^{-1} u_x \right] , \notag \\
H = \int dx \left( - \frac{1}{2} u_x^2 \right) , \label{e40}
\end{gather}
where $D$ and $H$ denote the Hamiltonian operator and functional, respectively, so that $u_t = D ( \delta H / \delta u )$ is the evolutionary form of \eqref{e3} for $D$ and $H$ given either by \eqref{e39} or by \eqref{e40}.

Finally, let us proceed to the soliton solutions of the SCPE \eqref{e3}. We derive them from the known soliton solutions of the sine-Gordon equation \eqref{e32}, using the transformation \eqref{e31}. For any given solution $s(y,t)$ of the sine-Gordon equation \eqref{e32}, the relations \eqref{e31} determine $u$ as a function of $y$ and $t$ uniquely, and determine $x$ as a function of $y$ and $t$ up to an additive constant of integration. This determines a solution $u(x,t)$ of the SCPE \eqref{e3}, given in a parametric form, with $y$ being the parameter. The invariance of the sine-Gordon equation \eqref{e32} under the Lorentz transformation
\begin{equation}
y \mapsto \gamma y , \qquad t \mapsto \gamma^{-1} t , \qquad s \mapsto s \label{e41}
\end{equation}
corresponds via \eqref{e31} to the invariance of the SCPE \eqref{e3} under the scale transformation
\begin{equation}
x \mapsto \gamma x , \qquad t \mapsto \gamma^{-1} t , \qquad u \mapsto \gamma u , \label{e42}
\end{equation}
where $\gamma$ is any nonzero constant. We can put the source solution of the sine-Gordon equation into a simpler form by \eqref{e41}, in order to simplify the symbolic integration required to obtain $x(y,t)$. Then we can use \eqref{e42} to generalize the target solution of the SCPE, if necessary. Also we can simplify the source solution of the sine-Gordon equation by shifts of $y$ and $t$, $y \mapsto y + y_0$ and $t \mapsto t + t_0$. A shift of $t$ in $s(y,t)$ causes the same shift of $t$ in $u(x,t)$, while a shift of $y$ has no effect on the target solution of the SCPE.

\begin{figure}
\includegraphics[width=12cm]{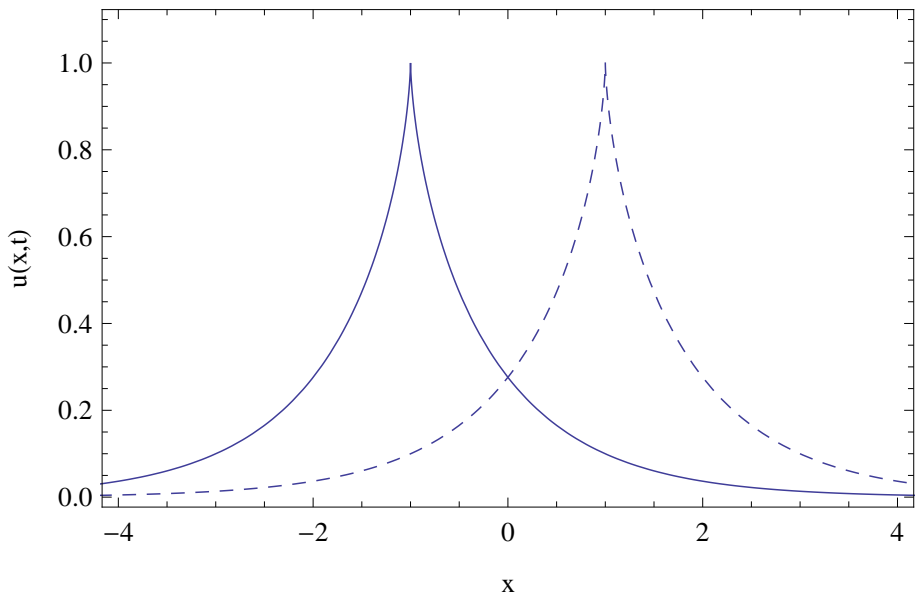}
\caption{The cusped soliton \eqref{e44}: $t = 1$ (solid) and $t = -1$ (dashed). \label{f1}}
\end{figure}

Taking the kink solution \cite{L} of the sine-Gordon equation \eqref{e32}, in the form
\begin{equation}
s = 4 \arctan [ \exp ( y + t ) ] \label{e43}
\end{equation}
simplified by the Lorentz transformation \eqref{e41} and a shift of $y$, we obtain via the transformation \eqref{e31} the following parametric expressions for the corresponding solution of the SCPE \eqref{e3}:
\begin{equation}
u = 1 / \cosh ( y + t ) , \qquad x = y - \tanh ( y + t ) , \label{e44}
\end{equation}
where $y$ serves as the parameter, $- \infty < y < \infty$, and the constant of integration in $x$ has been fixed so that $x |_{y=t=0} = 0$. (If we took the antikink solution of the sine-Gordon equation as a source solution for the transformation \eqref{e31}, the target solution of the SCPE would differ from \eqref{e44} in the sign of $u$ only.) This solution \eqref{e44} is the cusped soliton (cuspon) shown in Figure~\ref{f1}, which moves from the right to the left with constant shape and unit speed.
The angle of the cusp of this soliton is zero, because the approximation
\begin{equation}
u \approx 1 - \frac{1}{\sqrt[3]{2}} ( x + t )^{2/3} \label{e45}
\end{equation}
is valid for $| x + t | \ll 1$. We can generalize this solution \eqref{e44} by the scale transformation \eqref{e42}, thus obtaining either a bigger and faster cuspon or a smaller and slower one. Let us also remind that the soliton solution of the SPE \eqref{e2}, which corresponds via the transformation \eqref{e28} to the same kink solution of the sine-Gordon equation, is the loop soliton \cite{SS2}.

It is easy to see why and when the transformation \eqref{e31}, being applied to a smooth solution of the sine-Gordon equation \eqref{e32}, generates a solution of the SCPE \eqref{e3} with a singularity. From \eqref{e31} we get the relation
\begin{equation}
u_x (x,t) = \frac{\sin s(y,t)}{1 + \cos s(y,t)} . \label{e46}
\end{equation}
Then it immediately follows from \eqref{e46} that the target solution $u(x,t)$ of the SCPE can be free from singularities only if the corresponding source solution $s(y,t)$ of the sine-Gordon equation nowhere reaches any of the values
\begin{equation}
s = \pi \pm 2 \pi k , \qquad k = 0,1,2, \dotsc , \label{e47}
\end{equation}
for which $| u_x | \to \infty$. Any solution of the sine-Gordon equation, which contains asymptotically free kinks or antikinks at large $t$, does not satisfy this requirement, and the corresponding solution of the SCPE has to contain cusps therefore. Consequently, in order to obtain any smooth solution of the SCPE, we have to take a source solution of the sine-Gordon equation containing only breathers, which are known to be the bound kink-antikink states.

Let us take the breather solution \cite{L} of the sine-Gordon equation \eqref{e32}, simplified by the Lorentz transformation \eqref{e41} and shifts of $y$ and $t$, that is
\begin{equation}
s = - 4 \arctan \left( \frac{m \sin \psi}{n \cosh \phi} \right) , \label{e48}
\end{equation}
where $m$ is a constant, $0 < m < 1$, and
\begin{equation}
n = \sqrt{1 - m^2} , \qquad \phi = m ( y + t ) , \qquad \psi = n ( y - t ) . \label{e49}
\end{equation}
Applying the transformation \eqref{e31} to the solution \eqref{e48}, we obtain the following parametric expressions for the corresponding solution of the SCPE \eqref{e3}:
\begin{gather}
u = 2 m n \frac{m \sin \psi \sinh \phi + n \cos \psi \cosh \phi}{m^2 \sin^2 \psi + n^2 \cosh^2 \phi} , \notag \\
x = y + m n \frac{m \sin 2 \psi - n \sinh 2 \phi}{m^2 \sin^2 \psi + n^2 \cosh^2 \phi} , \label{e50}
\end{gather}
where $y$ serves as the parameter, $- \infty < y < \infty$, and the constant of integration in $x$ has been fixed so that $x |_{y=t=0} = 0$. Of course, this solution \eqref{e50} can be generalized by the scale transformation \eqref{e42} and shifts of $x$ and $t$, if necessary.

\begin{figure}
\includegraphics[width=12cm]{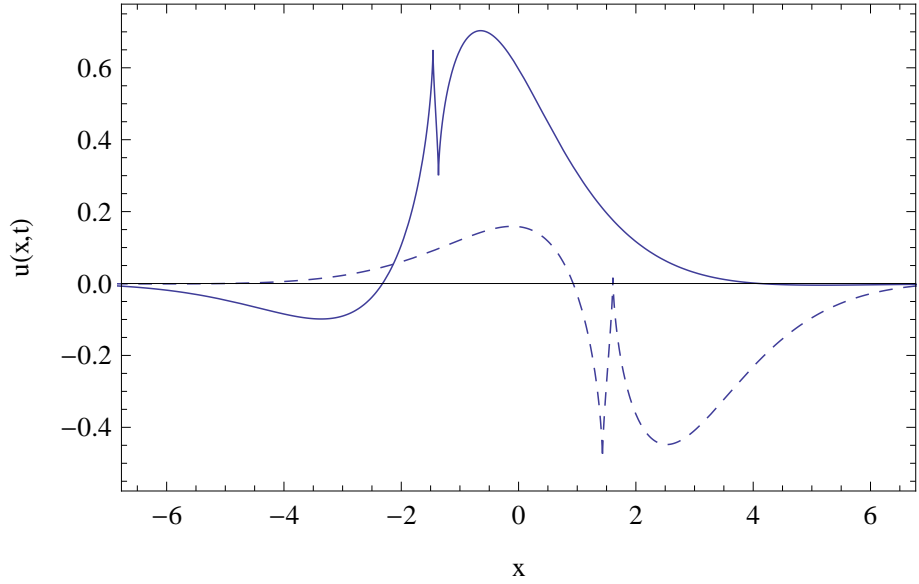}
\caption{The overcritical envelope soliton \eqref{e50} with $m = 0.85 > m_{\text{cr}}$: $t = 0.9$ (solid) and $t = -1.8$ (dashed). \label{f2}}
\end{figure}

\begin{figure}
\includegraphics[width=12cm]{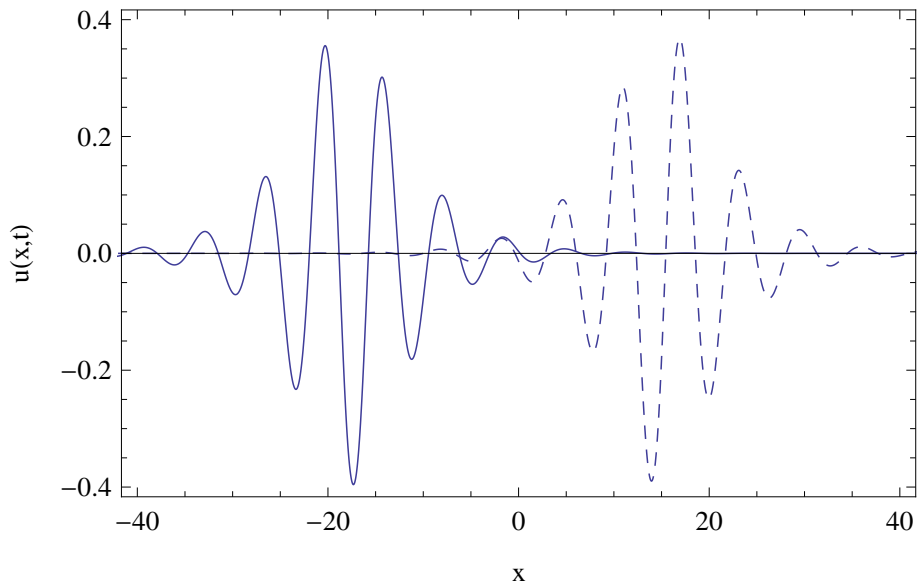}
\caption{The undercritical envelope soliton \eqref{e50} with $m = 0.2 < m_{\text{cr}}$: $t = 15$ (solid) and $t = -15$ (dashed). \label{f3}}
\end{figure}

\begin{figure}
\includegraphics[width=12cm]{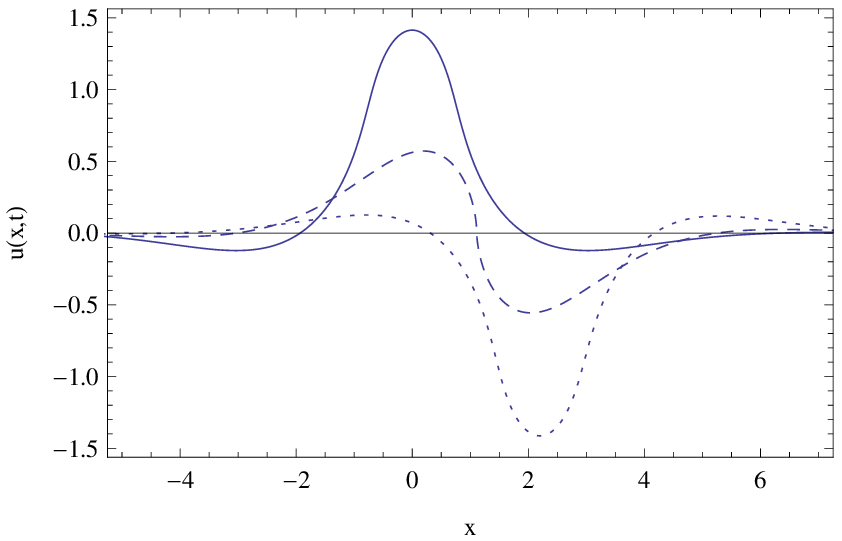}
\caption{The critical envelope soliton \eqref{e50} with $m = 0.707 \approx m_{\text{cr}}$: $t = 0$ (solid), $t = -1.1$ (dashed), and $t = -2.2$ (dotted). \label{f4}}
\end{figure}

The obtained solution \eqref{e50} of the SCPE \eqref{e3}, which we call the pulse solution or the envelope soliton, is free from singularities not for all values of $m$. It is easy to see that the function $s(y,t)$ given by \eqref{e48} does not reach any of the values listed in \eqref{e47}, for all $y$ and $t$, if and only if
\begin{equation}
0 < m < m_{\text{cr}} = 1 / \sqrt{2} \approx 0.707 . \label{e51}
\end{equation}
Therefore, in the overcritical case, when $m > m_{\text{cr}}$, the pulse solution \eqref{e50} contains cusps, as shown in Figure~\ref{f2}.
In the undercritical case, when $m < m_{\text{cr}}$, the pulse solution \eqref{e50} represents a smooth envelope soliton, a typical example of which with a small value of $m$ is shown in Figure~\ref{f3}.
The envelope curve of this pulse is determined by the hyperbolic functions in \eqref{e50} and moves from the right to the left. The oscillatory component of this pulse is determined by the trigonometric functions in \eqref{e50} and moves from the left to the right. The smaller the value of $m$, the larger the number of oscillations in the pulse. For very small values of $m$, $m \ll 1$, we find from \eqref{e50} that $x \approx y$ and
\begin{equation}
u \approx 2 m \frac{\cos ( x - t )}{\cosh [ m ( x + t ) ]} . \label{e52}
\end{equation}
On the other hand, if the value of $m$ tends to $m_{\text{cr}}$ in the undercritical case, the smooth envelope soliton of the SCPE \eqref{e3} can be as short as only one cycle of its carrier frequency. This is shown in Figure~\ref{f4}.
Let us also remind that the smooth envelope soliton of the SPE \eqref{e2}, which corresponds via the transformation \eqref{e28} to the same breather solution of the sine-Gordon equation, cannot be shorter than approximately three cycles of its carrier frequency \cite{SS2}.

\section{Conclusion} \label{s4}

In this paper, we have studied the integrability of a nonlinear wave equation which slightly generalizes the well-known integrable short pulse equation (SPE). We have transformed this generalized SPE to nonlinear Klein--Gordon equations whose nonlinearities depend on the coefficients of the generalized SPE. We have shown in this way that the generalized SPE is integrable in two distinct cases of its coefficients and that no more integrable cases should be expected unless the known list of integrable nonlinear Klein--Gordon equations is incomplete.

The first integrable case is the original SPE, while the second one is a new equation which we have called the single-cycle pulse equation (SCPE) due to properties of its solutions. The SPE and the SCPE are two different ``avatars'' of one and the same sine-Gordon equation. Moreover, the SCPE is a previously overlooked scalar reduction of the integrable system of coupled SPEs of Feng. We have obtained the Lax pair and bi-Hamiltonian structure for the SCPE. From the kink and breather solutions of the sine-Gordon equation we have derived the corresponding cusped soliton and envelope soliton solutions of the SCPE. We have shown that the smooth envelope soliton of the SCPE can be as short as only one cycle of its carrier frequency.

Consequently, the SCPE is an interesting new equation of soliton theory, which deserves further investigation in many aspect, including its hierarchy, conserved quantities, multi-soliton and periodic solutions, problems of wave breaking and well-posedness, integrable discretizations and multi-component generalizations. Moreover, owing to the properties of its smooth envelope soliton, the SCPE is able to appear in physics and technology as a model equation describing the propagation of extremely short wave packets in certain media with cubic nonlinearities.

\section*{Acknowledgment}

The author is deeply grateful to the Max Planck Institute for Mathematics, where a part of this work was done, for hospitality and support.


\begin{thebibliography}{24}

\bibitem{BRT} R. Beals, M. Rabelo, K. Tenenblat, B\"{a}cklund transformations and inverse scattering solutions for some pseudospherical surface equations, Stud. Appl. Math. 81 (1989) 125--151.

\bibitem{R} M.L. Rabelo, On equations which describe pseudospherical surfaces, Stud. Appl. Math. 81 (1989) 221--248.

\bibitem{SW} T. Sch\"{a}fer, C.E. Wayne, Propagation of ultra-short optical pulses in cubic nonlinear media, Physica D 196 (2004) 90--105.

\bibitem{CJSW} Y. Chung, C.K.R.T. Jones, T. Sch\"{a}fer, C.E. Wayne, Ultra-short pulses in linear and nonlinear media, Nonlinearity 18 (2005) 1351--1374; arXiv: nlin/0408020.

\bibitem{SS1} A. Sakovich, S. Sakovich, The short pulse equation is integrable, J. Phys. Soc. Jpn. 74 (2005) 239--241; arXiv:nlin/0409034.

\bibitem{SS2} A. Sakovich, S. Sakovich, Solitary wave solutions of the short pulse equation, J. Phys. A 39 (2006) L361--L367; arXiv:nlin/0601019.

\bibitem{SS3} A. Sakovich, S. Sakovich, On transformations of the Rabelo equations, SIGMA 3 (2007) 086; arXiv:0705.2889.

\bibitem{B1} J.C. Brunelli, The short pulse hierarchy, J. Math. Phys. 46 (2005) 123507; arXiv:nlin/0601015.

\bibitem{B2} J.C. Brunelli, The bi-Hamiltonian structure of the short pulse equation, Phys. Lett. A 353 (2006) 475--478; arXiv:nlin/0601014.

\bibitem{M1} Y. Matsuno, Multiloop soliton and multibreather solutions of the short pulse model equation, J. Phys. Soc. Jpn. 76 (2007) 084003.

\bibitem{M2} Y. Matsuno, Periodic solutions of the short pulse model equation, J. Math. Phys. 49 (2008) 073508.

\bibitem{P1} E.J. Parkes, Some periodic and solitary travelling-wave solutions of the short-pulse equation, Chaos Solitons Fractals 38 (2008) 154--159.

\bibitem{P2} E.J. Parkes, A note on loop-soliton solutions of the short-pulse equation, Phys. Lett. A 374 (2010) 4321--4323.

\bibitem{LPS} Y. Liu, D. Pelinovsky, A. Sakovich, Wave breaking in the short-pulse equation, Dyn. Part. Diff. Eqs. 6 (2009) 291--310; arXiv:0905.4668.

\bibitem{PS} D. Pelinovsky, A. Sakovich, Global well-posedness of the short-pulse and sine-Gordon equations in energy space, Commun. Part. Diff. Eqs. 35 (2010)  613--629; arXiv:0809.5052.

\bibitem{FMO} B.F. Feng, K. Maruno, Y. Ohta, Integrable discretizations of the short pulse equation, J. Phys. A 43 (2010) 085203; arXiv:0912.1914.

\bibitem{ZS} A.V. Zhiber, A.B. Shabat, Klein--Gordon equations with a nontrivial group, Sov. Phys. Dokl. 24 (1979) 607--609.

\bibitem{S1} S.Yu. Sakovich, On integrability of one third-order nonlinear evolution equation, Phys. Lett. A 314 (2003) 232--238; arXiv:nlin/0303040.

\bibitem{S2} S. Sakovich, Smooth soliton solutions of a new integrable equation by Qiao, J. Math. Phys. 52 (2011) 023509; arXiv:1010.1907.

\bibitem{BS1} J.C. Brunelli, S. Sakovich, On integrability of the Yao--Zeng two-component short-pulse equation, Phys. Lett. A 377 (2012) 80--82; arXiv:1205.6969.

\bibitem{BS2} J.C. Brunelli, S. Sakovich, Hamiltonian structures for the Ostrovsky--Vakhnenko equation, Commun. Nonlinear Sci. Numer. Simul. 18 (2013) 56--62; arXiv:1202.5129.

\bibitem{BS3} J.C. Brunelli, S. Sakovich, Hamiltonian integrability of two-component short pulse equations, J. Math. Phys. 54 (2013) 012701; arXiv:1210.5265.

\bibitem{F} B.F. Feng, An integrable coupled short pulse equation, J. Phys. A 45 (2012) 085202.

\bibitem{L} G.L. Lamb, Jr., Elements of Soliton Theory, Wiley, New York, 1980. 

\end{thebibliography}
\end{document}